\documentclass[twocolumn,twoside,slac_two]{revtex4}
\usepackage{graphicx}
\usepackage{fancyhdr}
\newcommand{\MSbar}{\overline{\rm MS}}

\begin{document}

\title{Theory review of  exclusive rare radiative decays}

%

\author{Ben D. Pecjak}
\affiliation{DESY theory group, Hamburg, Germany}

\begin{abstract}
I briefly review the theory status of exclusive rare radiative decays.
\end{abstract}

\maketitle

\thispagestyle{fancy}


\section{Introduction}
Experimental measurements of $B\to V\gamma$ decays, 
with $V$ a light vector meson such as $K^*,\rho,\omega,\phi$, 
have continued to improve and will become more precise 
at the end of the $B$-factories and at LHC-b.  Given their 
rich CKM phenomenology and  potential to constrain
new physics models, having reliable theory predictions for these 
decays is increasingly relevant.  In this talk I briefly
review the status of this area. 

The starting point is the 
effective weak Hamiltonian, which 
for $B\to V\gamma$ 
decays is~\cite{Buchalla:1996vs}: 
\begin{equation}
\mathcal{H}_{\rm eff} = 
\frac{G_F}{\sqrt 2} \sum_{p=u,c} \lambda_p^{(q)}  
\left [ C_1  Q_1^p + C_2  Q_2^p + 
\sum_{i=3}^8 C_i Q_i \right ] , 
\label{eq:heff}
\end{equation}
where $\lambda_p^{(q)} = V^*_{pq} V_{pb}$. 
The operators with the largest Wilson coefficients
are the four-quark operators 
$Q_1^p$ and $Q_2^p$, which read
\begin{eqnarray}
\label{eq:4-quark-operators}
 Q_1^p & =& (\bar q \, p)_{V-A} \, (\bar p \, b)_{V-A} \nonumber \\
 Q_2^p &=& (\bar q_i p_j)_{V-A} \,(\bar p_j b_i)_{V-A}, 
\end{eqnarray}
and the electromagnetic and chromomagnetic penguin 
operators $Q_7$ and $ Q_8$, which are
\begin{eqnarray}
 Q_7 &=& -\frac{e \, \overline m_b(\mu)}{8\pi^2}\,
\bar q \, \sigma^{\mu\nu} \, [1 + \gamma_5] \, b 
 F_{\mu\nu} \, , \nonumber \\ 
 Q_8 &=& -\frac{g \, \overline m_b(\mu)}{8\pi^2}\,
\bar q \,\sigma^{\mu\nu} \, [1 + \gamma_5] \, T^a \, b 
 G^a_{\mu\nu} . 
\label{eq:penguin-operators} 
\end{eqnarray}
Here $q = d$ or~$s$, and the convention for the sign of the 
couplings corresponds to the covariant derivative
$iD_\mu = i\partial_\mu +  e Q_f A_\mu +  g T^a A_\mu^a$, 
with~$A_\mu$ and~$A_\mu^a$ representing the photon and gluon 
fields respectively, and $Q_e = -1$ etc.  The factor
$\overline m_b(\mu)$ is the $\MSbar$ mass of the $b$ quark.

The main theoretical challenge is 
to evaluate the hadronic matrix elements of the operators
in the effective weak Hamiltonian.  Common 
ways of doing this include the QCD factorization 
\cite{Beneke:2001at, Bosch:2001gv, Ali:2001ez}
and pQCD \cite{Keum:2004is, Lu:2005yz} approaches.  Both 
of these rely on the fact that $\alpha_s(m_b)$ and $\Lambda_{\rm QCD}/m_b$
can be considered as small expansion parameters.
In this talk I will focus exclusively on QCD factorization 
methods, including in this category also strategies 
which supply additional information on $1/m_b$-suppressed
contributions from QCD sum rules \cite{Ball:2006nr, Ball:2006eu}, or 
apply a form of renormalization-group (RG) improved perturbation theory
based on soft-collinear effective theory (SCET) 
\cite{Becher:2005fg,Ali:2007sj}.

The remainder of the talk is organized as follows.  Section \ref{sec:QCDF}
introduces the QCD factorization formalism, Section \ref{sec:applications}
gives two sample applications, and Section \ref{sec:NNLO} reviews
recent results in higher-order perturbative corrections to the
hard-scattering kernels.  This is followed by a short discussion
of $1/m_b$ power corrections and endpoint divergences 
in Section \ref{sec:isospin}, using
isospin violation in $B\to K^*\gamma$ decays as an example, and 
a summary in Section \ref{sec:summary}.

\section{QCD factorization }
\label{sec:QCDF}
QCD factorization is the statement that in the heavy-quark limit
the hadronic matrix element of each operator in the 
effective weak Hamiltonian can be written in the form 
\begin{eqnarray} 
\label{eq:ff} 
&&\left \langle V \gamma \left | Q_i 
\right | \bar B \right \rangle = \zeta_{V_\perp}  \, t_i^{\rm I} + \\
&&
\frac{\sqrt{m_B}F f_{V_\perp}}{4}\phi^B_+
\otimes t^{\rm II}_i 
\otimes \phi^V_\perp  \, 
 \, 
+{\cal O}\left(\frac{\Lambda_{\rm QCD}}{m_b}\right), \nonumber
\end{eqnarray}
where the $\otimes$ stand for convolution integrals.  
(Sometimes this same formula is 
written in a different but equivalent form, using the tensor QCD form
factor $T_1^{B\to V}$ instead of the SCET soft function $\zeta_{V_\perp}$.)
The soft function $\zeta_{V_\perp}$, the
meson decay constants $f_{V_\perp},\, F$, and the light-cone
distribution amplitudes (LCDAs) $\phi^B_+, \, \phi^V_\perp$  are
non-perturbative but universal objects.  At present, numerical
values for these objects are taken from light-cone sum rules.
The hard-scattering kernels $t_i^{\rm I, II}$ can be calculated 
as a perturbative series
in $\alpha_s$.  The $t_i^{\rm I}$ are referred to as 
``vertex corrections'',  and the $t_i^{\rm II}$ are referred 
to as ``spectator corrections''.
Only the hard-scattering $t_7^{\rm I}$ is non-vanishing
at zeroth order in $\alpha_s$.  Corrections from this and 
the other operators appear at order $\alpha_s$, and
 have been known completely for some time 
\cite{Beneke:2001at, Bosch:2001gv, Ali:2001ez}.
I shall refer to these $\alpha_s$ corrections as next-to-leading 
order (NLO), even though the hard-spectator kernels $t_i^{\rm II}$
first start at this order.  In Section \ref{sec:NNLO} I describe
some recent results for the NNLO kernels.

An all orders proof of the QCD factorization
formula (\ref{eq:ff}) was performed in \cite{Becher:2005fg}, using
the technology of SCET.  In the effective-theory approach the 
hard-scattering kernels are short distance Wilson coefficients
of operators whose hadronic matrix elements define the $\zeta_{V_\perp}$
and the LCDAs.  For the vertex term this makes little practical
difference, but for the hard-spectator term one can show that 
the hard-scattering kernels $t_i^{\rm II}$ can be
further factorized into the form
\begin{equation}\label{eq:CJ}
t_i^{{ \rm II}}(u,\omega,\mu_i)= [C_i^{B1}(\mu_h) \otimes U(\mu_h,\mu_i)]\otimes 
j_\perp(\mu_i)\,.
\end{equation} 
The hard coefficients $C_i^{B1}$ contain physics at the scale $m_b$,
while the jet function $j_\perp$ is independent of the operator
and contains physics at the intermediate 
scale $\sqrt{m_b \Lambda_{\rm QCD}}\sim 1.5\,{\rm GeV}$. 
The evolution factor $U$ is derived
by solving the RG-equations in the effective theory. Because these
equations are non-local, the evolution factor appears in 
a convolution with the Wilson coefficient $C_i^{B1}$.  In the 
limit where  $m_b^2\gg m_b\Lambda_{\rm QCD}$, this evolution matrix
resums perturbative logs in the ratio $\Lambda_{\rm QCD}/m_b$,
and allows one to evaluate the hard coefficients and the jet function
at their natural scales $\mu_h\sim m_b$ and 
$\mu_i\sim \sqrt{m_b\Lambda_{\rm QCD}}$.  If one does not wish to do
this, then making the choice $\mu_h=\mu_i$ sets the evolution factor
$U$ to unity, and one recovers the original
QCD factorization formula (\ref{eq:ff}). So for these decays,
the ``SCET approach'' is just QCD factorization, plus the added
opportunity to perform a further scale separation in the hard
scattering kernels $t^{\rm II}_i$. 

It is necessary to keep in mind that the factorization formula
(\ref{eq:ff}) is valid only up to  
${\cal O}(\Lambda_{\rm QCD}/m_b)$ power corrections.  
Since SCET is an effective
theory which sets up a systematic expansion in $\alpha_s$ and 
$\Lambda_{\rm QCD}/m_b$, it has the potential to provide a complete
classification of the subleading terms in $1/m_b$.  
However, there has been no serious attempt to do this, 
because the generalization of the leading-order factorization 
formula to corrections in $1/m_b$  presumably contains many more 
non-perturbative objects. Even more 
troublesome is the fact that, in the cases where power 
corrections been calculated, the convolution integrals over 
momentum fractions do not always converge.  These ``endpoint divergences''
are at present a principle limitation on the entire formalism. 

In the absence of a comprehensive theoretical framework to deal
with the $1/m_b$ suppressed corrections, it has become common
practice to focus on the contributions which are believed
to be large, or which play an especially important role for
phenomenology.  In this talk I will discuss  the 
power corrections which are important for the calculation 
of isospin violation in $B\to K^*\gamma$ decays as performed
in \cite{Kagan:2001zk}.  This is sufficient to describe the
problem of endpoint divergences.  
Other power corrections, due to long-distance photon emission and soft-gluon
emission from quark loops, were calculated within the context of light-cone
sum rules in  \cite{Ball:2006eu}.

\section{Sample applications}
\label{sec:applications}
The formalism above can be used to calculate many different
observables, see for instance 
\cite{Bosch:2004nd, Beneke:2004dp, Ali:2006fa}.
In this section I briefly touch on two examples, taking 
results from the recent studies  \cite{Ball:2006nr, Ball:2006eu}  
for simplicity.  The first is the determination of the ratio of CKM
matrix elements $|V_{td}/V_{ts}|$ from the ratio of branching
fractions in $B\to\rho\gamma$ and $B\to K^* \gamma$ decays.
This is an independent check on the measurements from $B_s$ oscillations
\cite{Abulencia:2006ze}.  The most recent theory results are
\cite{Ball:2006nr}
\begin{eqnarray}
\label{eq:R}
R &\equiv&\frac{\overline{\cal B}(B\to (\rho,\omega)\gamma)}
{\overline{\cal B}(B\to K^*\gamma)} = \frac{|V_{td}|^2}{|V_{ts}|^2} \\
&\times& \left(0.75 \pm 0.11(\xi)\pm 0.02(\mbox{UT param., ${\cal O}(1/m_b)$})\right) \,,
\nonumber 
\end{eqnarray}
which combined with February 2007 HFAG data leads 
to $|V_{td}|/|V_{ts}|=0.192 \pm 0.014 ({\rm th})\pm 0.016({\rm exp})$.
This central value is compatible with that from $B_s$ oscillations, 
although the experimental errors are significantly larger and the theory 
errors about twice as large. The dominant theory error is in the 
form factor ratio $\xi= T_1^{B\to K^*}/T_1^{B\to \rho}$, and according
to the recent review \cite{Buchalla:2008jp} this should go down 
by about a factor of two with improved lattice results for the tensor decay
constants.  It should be mentioned that the experimental 
measurements of the ratio in (\ref{eq:R}) combine data 
from $\rho$ and $\omega$ decays under the assumption of exact isospin 
asymmetry, while the theory calculations can actually predict the
magnitude of isospin breaking.  Improved experimental 
measurements such as those in \cite{Aubert:2006pu, Taniguchi:2008ty}
will make this isospin averaging unnecessary.

As a second application I quote results for the isospin asymmetry in
$B\to K^*\gamma$ decays.  This asymmetry was calculated in the context of
QCD factorization in  \cite{Kagan:2001zk} and the numbers updated
in  \cite{Ball:2006eu}, which gives the result
\begin{eqnarray}
A_I(K^*)&=&\frac{\Gamma(\bar B^0\to \bar K^{*0}\gamma)
-\Gamma(B^- \to \bar K^{*-}\gamma)}
{\Gamma(\bar B^0\to \bar K^{*0}\gamma) 
+\Gamma(B^- \to \bar K^{*-}\gamma)} \nonumber \\
&=&(5.4\pm 1.4)\%.
\end{eqnarray}
The current result by HFAG in \cite{Barberio:2007cr} is 
$(3\pm 4)\%$.  Results from \cite{Kagan:2001zk} showed that this asymmetry
is particularly sensitive to the penguin operator $Q_6$.  
Improved experimental measurements of this asymmetry can 
thus provide constraints on new physics.  The theory prediction 
above, however, does not include any $\alpha_s$ corrections
to the perturbative hard-scattering kernels, even though some 
of these were calculated in \cite{Kagan:2001zk}.  The reason for this
is the presence of endpoint divergences, a topic discussed
in more detail in Section \ref{sec:isospin}.

\section{NNLO perturbative corrections}
\label{sec:NNLO}
Recently, a set of the NNLO corrections to the vertex and 
spectator kernels were obtained in \cite{Ali:2007sj}.  
In this section I summarize the new results from that work
and also what remains to be done to complete the NNLO calculation.

Consider first the vertex corrections and the hard-scattering
kernel $t^{\rm I}$. This hard-scattering kernel is most 
easily obtained as a matching coefficient in SCET;  
details were given in  \cite{Becher:2005fg, Ali:2007sj}.
The only technical point I mention here is that the 
matching coefficients are independent of the external 
states used in the calculation, and it is possible to use
the partonic matrix element $\langle q\gamma|Q_i|b\rangle$ 
in the matching.  The SCET calculation is trivial for 
on-shell quarks, because the loop corrections are given by 
scaleless integrals which vanish in dimensional regularization, 
so the main challenge is to calculate the partonic matrix elements
in full QCD.  However, these are just virtual corrections to
the $b\to q\gamma$ process needed also for inclusive 
$B\to X_s\gamma$ decay, and are currently
known very accurately due to the efforts of many people (see 
\cite{ Ali:2007sj, Misiak:2006zs} for a complete list of references).  Using
those multi-loop calculations, $t_{7,8}^{\rm II}$ can be obtained
completely to NNLO.  For the four-quark operators $Q_{1,2}$, on
the other hand, results to NNLO are known only 
in the large-$\beta_0$ limit, obtained by calculating
the $\alpha_s^2 C_F n_f$ terms and replacing $n_f\to -3/2\beta_0$, 
where $\beta_0=11 - 2/3 n_f$ in QCD. The remaining
terms require to calculate a large number of three-loop graphs
depending on the ratio $m_c^2/m_b^2$ and containing imaginary parts.
Given that this is also a missing piece in the $B\to X_s\gamma$
calculation, it is likely that this will be done in the near 
future.  

I now turn to the spectator corrections and the hard scattering
kernels $t_i^{\rm II}$.  Again in this case their calculation
is most conveniently formulated as a matching calculation
in SCET.  In contrast to the vertex corrections, however,
the Feynman diagrams needed in the matching calculation
have no analog in inclusive decay, where interactions with
the spectator quarks do not contribute at leading power in 
$1/m_b$.  Therefore, results for the $t_i^{\rm II}$ cannot simply
be extracted from existing calculations and must be obtained
from scratch.  The one exception is 
the $\alpha_s^2$ contribution to $t_7^{\rm II}$,
which can be taken from studies of heavy-to-light form 
factors carried out in \cite{Beneke:2004rc, Becher:2004kk}. 
The $\alpha_s^2$ contribution to $t_8^{\rm II}$ was calculated
in \cite{Ali:2007sj}.  The most important missing pieces 
are from $Q_{1,2}$, but these are also the hardest to calculate,
since they involve two-loop graphs depending on $m_c^2/m_b^2$
in addition to the momentum fraction $u$ of the quark in the 
$V$-meson.  

A natural place to explore the numerical impact of the 
NNLO corrections is on branching fractions in  
$B\to K^*\gamma$ decays, since for these
decays the annihilation topology is CKM suppressed.  
The branching fraction is 
\begin{equation}
\label{eq:BF}
{\cal B}(B\to K^*\gamma)=\frac{\tau_B m_B}{4\pi}\left(1-
\frac{m_{K^*}^2}{m_B^2}\right)\left | {\cal A}_{\rm v}+ {\cal A}_{\rm hs}
\right|^2\, ,
\end{equation}
where the ${\cal A}_{\rm v}$ (${\cal A}_{\rm hs}$) are the amplitudes
for the vertex (hard spectator) corrections.  These are separately
RG-invariant and so can be studied individually.  For the vertex
corrections, the ratio of the NNLO amplitude to LO amplitude for
the default set of parameters in \cite{Ali:2007sj} is
\begin{eqnarray}
\frac{{\cal A}_{\rm v}^{\rm NNLO}}{{\cal A}_{\rm v}^{\rm LO}}
= 1&+&\left(0.096+0.057 i\right)\left[\alpha_s\right] \nonumber \\
&+& \left(-0.007+0.030 i\right)\left[\alpha_s^2\right], \nonumber
\end{eqnarray}
where the first term in parentheses is the NLO ($\alpha_s)$
correction and the second term the NNLO  $(\alpha_s^2)$ 
correction.  One sees that  the real part of the NNLO correction is 
extremely small.  It is instructive to further split the 
above amplitude into the pieces originating from the 
various operators $Q_i$.  In that case the result reads
{\small \begin{eqnarray}\label{eq:split}
&&\frac{{\cal A}_{\rm v}^{\rm NNLO}}{{\cal A}_{\rm v}^{\rm LO}}-1= \\
&&\bigg((.26+  .03i)\, [Q_1] -.18 \,[Q_7] 
+(.02+.02i) \,[ Q_8] \bigg)\,[\alpha_s]
\nonumber \\
&&+ \bigg((.07+.02i)\, [Q_1] -.08 \,[Q_7] 
+(.002+.01i) \,[ Q_8] \bigg)\,[\alpha_s^2] \nonumber \,.
\end{eqnarray}}
From this one sees that the NNLO contributions are so small
because of a large cancellation between the $Q_1$ and 
$Q_7$ contributions. It is an open question whether 
this cancellation will persist when results for $Q_1$ are 
obtained beyond the large-$\beta_0$ approximation used above.
This will be discussed in more detail below, where results 
for branching fractions are given.    
 
For the hard spectator amplitude, one finds 
\begin{equation}
\frac{ {\cal A}_{\rm hs}^{\rm NNLO}}{{\cal A}_{\rm v}^{\rm LO}}=
\big(0.11+0.05 i\big)\,[\alpha_s]+
\big(0.03+0.01 i\big)\,[\alpha_s^2] .
\end{equation}
Unlike the case of the vertex corrections,
the individual contributions from the different operators 
are rather small at NLO and especially NNLO.  The exact 
numbers, including a split into components at the jet 
and hard scales in (\ref{eq:CJ}), can be found in 
\cite{Ali:2007sj}. 

Putting the amplitudes together, one can find the branching
fractions and their uncertainties.  Including isospin 
and SU(3) breaking from meson masses, lifetimes, and
$V$-meson distribution amplitudes, \cite{Ali:2007sj} 
estimated
\begin{eqnarray}
{\cal B}(B^+ \to K^{*+}\gamma) &=& 
(4.6 \pm 1.4 ) \times 10^{-5} ,\nonumber\\
{\cal B}(B^0 \to K^{*0}\gamma) &=& 
(4.3 \pm 1.4)  
\times 10^{-5} ,\nonumber\\
{\cal B}(B_s \to \phi\gamma) &=& (4.3
\pm 1.4) \times 10^{-5}.
\label{eq:finalbrs}
\end{eqnarray}  
The magnitude of various sources of uncertainty can 
be found in \cite{Ali:2007sj}.  Most significant
is about a 25\% uncertainty due to the soft functions
$\zeta_{V_\perp}$, which is  expected.
Somewhat surprising is that the $\alpha_s^2$ corrections
from $Q_1$ can make a large impact, and since they 
are taken only in the large-$\beta_0$ limit the true
result is still rather uncertain.  For instance, 
assigning a $100\%$ uncertainty to the 
$\alpha_s^2$ piece from $Q_1$ in (\ref{eq:split}),
one finds an uncertainty of about $\pm 0.5$ in the 
branching fractions.  Also, because the $C_F n_f$ terms
do not fix the perturbative definition of the 
charm-quark mass in the $\alpha_s$ contribution in (\ref{eq:split}),
one can formally use values ranging from a low $\MSbar$ mass to 
a high pole mass, which can make a large numerical difference,
about $\pm 0.4$ in the range used in  \cite{Ali:2007sj}.
The message to be gained from this is that the full NNLO corrections
from the set $Q_{1,2}$ also need to be calculated, in order to 
get the reduced perturbative uncertainty expected from 
a higher-order calculation.

\section{Isospin violation and endpoint divergences}
\label{sec:isospin}
\begin{figure}[t]
\centering
\includegraphics[width=80mm]{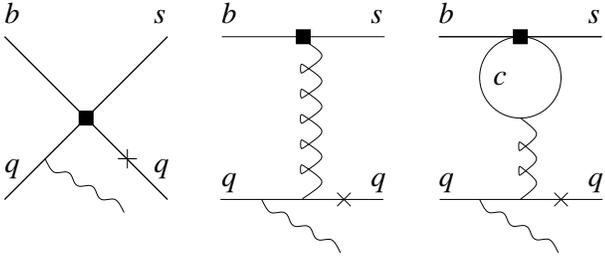}
\caption{Feynman diagrams contributing to isospin violation 
in $B\to K^*\gamma$ decays.} \label{fig:ann}
\end{figure}
Calculations of interesting observables such
as isospin violation in $B\to K^* \gamma$ decays require 
to include $1/m_b$ suppressed contributions to the 
factorization formula.  Particularly important 
corrections come from  topologies such as 
those in  Figure \ref{fig:ann}, which I have 
taken from \cite{Kagan:2001zk}. The square vertex
indicates an insertion of $Q_1\dots Q_6$ for the graph on the left,
an insertion of $Q_8$ for the graph in the center, and an 
insertion of $Q_{1,2}$ for the graph on the right.  The crosses
denote alternate attachments of the photon line. 
Note that the center and right graphs are $\alpha_s$ 
corrections compared to the graphs on the left.  
The contributions from the graphs in the figure, as well
as some $\alpha_s$ terms linked to the graph on the left
by RG-invariance, were calculated in \cite{Kagan:2001zk}.
To finish the calculation to this order would require to 
calculate the $\alpha_s$ corrections to the graphs on the left.
This calculation is in progress \cite{ongoing}.  However, 
once completed, the calculation will still face the problem
that a straightforward application of the QCD factorization formalism
leads to endpoint divergences in the convolution integrals.  For 
example, the center graph in Figure \ref{fig:ann} was shown in 
\cite{Kagan:2001zk} to contain the integral
\begin{equation}
\label{eq:div}
X_\perp = \int_0^1 du \, \phi_\perp^{K^*}(u)\frac{2-u}{3(1-u)^2},
\end{equation}
which is equal to infinity under the conventional assumption
that the LCDA vanishes as $1-u$ in the endpoint. The quantity
$X_\perp$ is multiplied by the Wilson coefficient $C_8$ and 
was estimated by introducing an IR cutoff in  \cite{Kagan:2001zk}
and found to be small, but clearly this is a conceptual problem.
Moreover, there is no guarantee that similar endpoint divergences
will not appear for the other operators in the weak Hamiltonian,
once the full set of one-loop corrections to the graphs on 
the left in Figure \ref{fig:ann} are included.  Therefore, it 
is fair to say that a better understanding of power corrections 
is needed, in order to achieve a consistent framework for the 
calculation of quantities such as isospin violation.

\section{Summary}
\label{sec:summary}
Rare radiative $B\to V\gamma$ decays are of increasing
interest as experimental measurements become more 
precise.  QCD factorization and SCET have provided a theoretical framework
which can be used to calculate observables for these decays to 
leading order in $1/m_b$.  The perturbative hard-scattering
kernels have been known at NLO in $\alpha_s$ for some time,
and recently a set of NNLO results have been obtained for the 
operators $Q_1$, $Q_7$, and $Q_8$, although results for the four-quark 
operators $Q_1$ and $Q_2$ are not yet complete.  The 
interesting observables such as isospin asymmetries and 
branching fractions in the $b\to d\gamma$ modes are sensitive
to power corrections in $1/m_b$.  Some of these have been estimated in the 
framework of light-cone sum rules, but a systematic treatment 
in SCET or QCD factorization is still missing due to the presence
of endpoint divergences; a solution to this problem would 
be a much desired advance in this field.

\begin{acknowledgments}
I would like to thank the organizers of the conference for the
invitation to give this talk.
\end{acknowledgments}

\bigskip 

\end{document}